\documentclass[12pt]{aastex}

\usepackage{color}
\usepackage{graphicx}
\usepackage{fancyheadings}
\usepackage{lscape}

\shorttitle{Chandra Observations of Arp 220}
\shortauthors{McDowell et al.}

\begin{document}


\title{ Chandra Observations of Extended X-ray Emission in Arp 220}

\author{J.C. McDowell}
\affil{Harvard-Smithsonian Centre for Astrophysics, 60 Garden Street,
Cambridge, MA 02138, USA}
\email{jcm@cfa.harvard.edu}

\author{D.L. Clements}
\affil{Physics Dept., Imperial College, Prince Consort Road, London SW7 2BW, UK}
\email{d.clements@ic.ac.uk} 

\author{S.A. Lamb}
\affil{Center for Theoretical Astrophysics, Departments of Physics and 
Astronomy, Loomis Laboratory, University of Illinois, 1110 W. Green 
Street, Urbana, IL 61801, USA}

\author{S. Shaked}
\affil{University of Arizona, Dept. Astronomy, 933 North Cherry Avenue,
Tucson, AZ 85721-0065, USA}

\author{N.C. Hearn}
\affil{Center for Theoretical Astrophysics, Department of Physics, Loomis 
Laboratory, University of Illinois, 1110 W. Green Street, Urbana, IL 
61801, USA}

\author{L. Colina}
\affil{lnstituto de Estructura de la Materia (CSIC), Serrano 121, 28006 Madrid,
Spain}

\author{C. Mundell}
\affil{ARI, Liverpool John Moores University, Twelve Quays House,
Egerton Warf, Birkenhead, Wirral, Cheshire, CH41 1LD, UK}

\author{K. Borne}
\affil{Raytheon Information Technology and Sciences Services, NASA Goddard
Space Flight Centre, Greenbelt, MD 20771, USA}

\author{A.C. Baker}
\affil{Dept. Physics and Astronomy, Cardiff University, PO Box 913, 
Cardiff,
CF24 3YB, UK}

\author{S. Arribas}
\affil{Space Telescope Science Institute,
ESA Space Telescope Division,
3700 San Martin Drive, Baltimore, MD 21218}


\begin{abstract}
\medskip

We resolve the extended X-ray emission from the prototypical
ultraluminous infrared galaxy Arp 220. Extended, faint edge-brightened,
soft X-ray lobes outside the optical galaxy are observed to a distance
of  10 to 15 kpc on each side of the nuclear region. Bright plumes
inside the optical isophotes coincide with the optical line emission and
extend 11 kpc from end to end across the nucleus. The data for the
plumes cannot be fit by a single temperature plasma, and display a range
of temperatures from 0.2 to 1 keV. The plumes emerge from bright,
diffuse circumnuclear emission in the inner 3 kpc centered on the
H$\alpha$ peak, which is displaced from the radio nuclei. There is a
close morphological correspondence between the H$\alpha$ and soft X-ray
emission on all spatial scales. We interpret the plumes as a starburst-driven
superwind, and discuss two interpretations of the emission from the
lobes in the context of simulations of the merger dynamics of Arp 220.

\smallskip
\end{abstract}


\section{Introduction}

The evolutionary importance of energetic events on galactic scales has
become central to our present understanding of cosmological evolution.
The realization in the IRAS era (e.g. Soifer et al. 1984) that 
super-starburst
ultraluminous infrared galaxies rivalled quasars in their energy output
made it clear that for at least a subset of galaxies  a large fraction
of their stellar content and gas mass is drastically rearranged during
interactions which last from $10^8$ to $10^9$ years (Clements et al. 
1996)
The observation of superwinds (Chevalier \& Clegg 1985)  emphasizes the role
of energetic events in galaxies in redistributing material both within a
galaxy and into the intergalactic medium. The energy and momentum
involved in the biggest train wrecks in the universe are released in a
number of different channels: stellar velocity dispersions, tidal
disruption, gas heating, mechanical energy deposited in the gas and
stars, shock-triggered star formation and subsequent infrared emission.


Truman H. Safford (1836-1901) discovered the nebula Safford 7 (IC 1127)
at Chicago  on 1866 May 4 (Safford 1887), but recorded its position with
an RA off by one minute of time. Javelle\footnote{Stephane Javelle
(1864-1917) is most famous for claiming to have seen a strange light on
Mars, and appears thinly disguised in H.G. Wells' {\it War of the Worlds} 
(1898)~as `Lavelle of Java'.} (1905) rediscovered it as Javelle 1368 at
Nice on 1903 Jul 25. It was recorded by Dreyer in IC2 (Dreyer 1910) as
IC 4553, but the object remained obscure until A.G Wilson identified it
on Palomar Sky Survey prints as unusual and Arp (1966) included it in
his catalog of peculiar galaxies as Arp 220, the designation by which it
is most commonly known today. The discovery of strong OH maser emission
(Baan, Wood \& Haschick 1982) and ultra-luminous far infrared output
(Soifer et al. 1984) established Arp 220 as the prototypical infrared
superluminous merger. ISO observations (e.g. Genzel et al. 1998, Lutz et 
al. 1998)
support a starburst interpretation for the bulk of the luminosity and
Scoville et al. (1998) locate a number of massive star clusters in the
nuclear region using NICMOS.
At a distance of 76 Mpc (z= 0.018; Kim \& Sanders 1998), it
is one of the most luminous objects in the local universe. The galaxy,
with an apparent B magnitude of 13.9 and a total infrared luminosity 
(10-100 microns)
of $1.5\times 10^{12} L_\odot$, is believed to have been formed by the
almost face-on collision of two spiral disks.

In the optical, Arp 220 appears irregular and dusty, but other wavebands
have been used to probe into the center of the object and reveal the
presence of activity within a few parsecs of the nucleus. VLBI 
observations
of Arp 220 suggest that the peak of the OH emission originates in a
structure $\lesssim$ 1 pc across (Lonsdale et al. 1994) and  Rieke et al.
(1985) claimed that as much as half of the luminosity in Arp 220 may 
be due to an active nucleus.  Armus et al. (1995) used near-infrared
spectra to suggest that as much as 80-90\% of the total luminosity is
powered by an obscured AGN.  Dudley and Wynn-Williams (1997) use the
depth of the silicate absorption feature to estimate a
size for the emission region at 10 microns of only a few parsecs.

Eales \& Arnaud (1988) reported the first detection of X-rays from Arp 
220,  using the Einstein IPC. However, their observation was confused by
emission from a neighbouring group of galaxies first noted by Heckman et
al. (1996) who carried out the first detailed study of the X-ray
properties of the galaxy with the ROSAT PSPC. Their data show that the
X-ray emission in the 0.1-2.4 keV band has a size of $\sim 30 \times 11$
kpc and a luminosity of between $4.3\times 10^{40}$ and $2.3\times
10^{41}$ ergs s$^{-1}$ and set an upper limit of $\sim 20\%$ of the
observed X-ray emission  associated with any single point source. They
noted that the spatial extent of the X-ray emission was around 22 kpc,
much larger than the IR-emitting region, and that the X-ray luminosity
was an order of magnitude larger than that of normal spiral galaxies.
The "double-bubble" morphology seen in optical emission-line images
(Heckman, Armus \& Miley 1987) coincided with the X-ray nebula seen with
ROSAT, but the spatial resolution of those data was insufficient for a
detailed  comparison. Heckman et al. suggested that both the X-ray and
line emission is due to a bipolar "superwind" driven out from the
nucleus by a starburst or a dust-shrouded QSO. Gallagher et al (2002)
report similar extended X-ray emission in the ultraluminous galaxy
Markarian 231, although at z=0.042 the spatial resolution of  their
Chandra image reveals less detail than the observations presented here.

\section{Observations and Data Analysis}

We observed Arp 220 on 2000 Jun 24 for 56 ks with the ACIS camera on the
Chandra X-ray Observatory as part of the Cycle 1 guest observer program.
Arp 220 was placed on the S3 back illuminated chip near the node 0/node
1 boundary. Here we report on the extended emission from Arp 220;
Clements et al (2002, Paper I) discuss the point sources in the nuclear
region. Data reduction was carried out with CIAO version 2.1 and 2.2.
The data were taken with the chip at a temperature of -120C and were
gain-corrected using the revised file acisD2000-01-29gainN0003.fits from 
the July 2001
recalibration, which improves the calibration at low energies. 
The observation was relatively unaffected by background
flares and only a small amount of exposure was removed, leaving  an
effective exposure time of 55756s. Astrometry was corrected using a
revised geometry file (telD1999-07-23geomN0004.fits) which is believed
to provide positions across the full ACIS field accurate to about 1
arcsecond.

The extended emission from Arp 220 is clearly visible in the raw data.
The west lobe is dithered partly across the node 0 boundary, but the
remainder of the emission is all on node 1. We generated spectral
responses making use of the recently recalibrated
(acisD2000-01-29fef\_piN0002.fits) low energy response, both using the
standard CIAO mkrmf tool and using the calcrmf tool developed by Alexey
Vikhlinin (available from the Chandra software swap page), which
produces a count-weighted average of the response. As expected, there
was very little difference in the results from the two methods.


For global images,  we used the background event files developed by
Markevitch et al (2000), together with the standard exposure map tools, to
generate background-subtracted, exposure corrected images in various
energy bands. These corrected images were then normalized to retain the
same mean number of counts, and adaptively smoothed using the csmooth
program. X-ray color images of the smoothed data (Figs. 1 to 4) show the
morphology of the region and allow regions of different spectral
properties to be distinguished. The smoothed images were used in further
analysis only to define extraction regions, not to determine any
numerical quantities.

For spectral studies we extracted PI count histograms for various
regions in the emission (Fig. 5). We made two background spectra, one
using the Markevitch background files and one using a local background
extracted from two 50-arcsecond-radius circles at either end of node 1
in our dataset and containing 10650 counts. The two spectra were
similar, and the results presented here use the local background. The
galactic extinction in the direction of Arp 220 has a column of
$4.1\times 10^{20} \mbox{cm}^{-2}$ (Stark et al. 1992) and this has been
used as a lower limit in our spectral fits.

\section{X-ray emission from Arp 220}

\subsection{Overall morphology}

The Chandra view of Arp 220 reveals structure on scales from one
arcsecond to several arcminutes. We can distinguish four scales of
interest: the nuclear region, the galaxy with two regions of extended
emission (`plumes'), the extended 20-kpc scale emission (`lobes'), and
the unrelated group of galaxies to the south reported by Heckman et al.
(1996). The position angles of the plumes and lobes are in agreement
with the PSPC and HRI structures described by Heckman et al.
Here we describe the observations, leaving most physical
interpretation to later sections.

The integrated spectrum from Arp 220 can be fit by the sum
of several thermal contributions and a power law, and gives good agreement
on the total flux with the individual fits to components, with 1694
net counts and a luminosity of $1.2\times 10^{41}$ erg s$^{-1}$,
in good agreement with the estimates of Heckman et al (1996).
The data on individual components are summarized in Table 1 and discussed
below.

\subsection{The circumnuclear region}

In the nuclear region we see hard emission on a scale of two to three
arcseconds, and indications of bright point sources, denoted X-1 and X-4
in Paper I (Clements et al. 2002) close to the nucleus. A soft emission
peak, denoted X-3 in Paper I, is also present and is extended over a
diameter of 2.5 kpc. Its centroid is displaced 1.5 arcseconds to the
northwest of the hard emission. The hard emission coincides with a dust
lane in the galaxy (Joy et al. 1986), and indeed the soft emission is
suppressed there. However, the absence of hard emission away from the
nucleus shows that the spectral change is due to a different type of
source, and not merely an absorption effect. In Paper I  we presented
reconstructed images and spectral fits to the nuclear sources X-1 to
X-4. 

\subsection{The plumes}

On a scale of a few kiloparsecs, we see two regions of bright extended
X-ray emission, one on either side of the nucleus, along a position
angle of 135-150 degrees. This emission was seen in the HRI observations
reported by Heckman et al. (1996); we will refer to these regions as the
NW and SE `plumes'. We cannot rule out that the emission comes
from unresolved point sources, but if the plumes are two unrelated
starburst regions we would expect binaries to contribute a hard
component to the spectrum.
Fig. 6 shows a contour plot of the reconstructed image of the plume region.

The bright region of the NW plume is about 3 kpc in extent  
and the SE plume is larger and
brighter, 3 x 5 kpc in extent; the projected tip-to-tip length of the
emission is 10 kpc. Each of the plumes appears to be roughly round, but with
an elongation in the radial direction for the SE plume. 
The emission is not sharp-edged, and the dimensions given here 
correspond to a contour of 10 percent of the peak surface brightness.

The east plume spectrum is inconsistent with that from 
a single temperature plasma. It
includes a soft thermal contribution with a temperature of less
than 0.25 keV, together with hotter thermal plasma or bremsstrahlung
ranging from 1 to 5 keV. There are fewer counts in the west plume,
but the spectrum definitely contains flux out to 2 keV. Formally, the 
data provides only an upper limit to the absorption column in the plumes of $1.5\times
10^{22}\mbox{cm$^{-2}$}$. The difference in inferred intrinsic
luminosity (source side of absorption) between assuming only foreground
absorption and using this upper limit renders estimates of the unabsorbed
luminosity for this soft emission uncertain by a factor of a thousand; the values
in Table 1 should therefore be considered illustrative.
The overall morphology of the plumes could also be affected by
patchy absorption in the galaxy, but the visible dust lanes do not correlate
with the plume boundaries except in the inner edge of the east plume.
Radio observations (Hibbard, Vacca and Yun 2000) show the presence of an
elongated HI emission feature to the NE and SW of the galaxy on 20 kpc
scales; notably, the HI emission avoids the
regions of X-ray emission - there is a gap in the HI which corresponds
to the area with the plumes and is aligned with them. However, the measured
column of the HI, less than $5\times 10^{19} \mbox{cm$^{-2}$}$ in the region
of the plumes, would provide too little absorption to affect the observed
soft X-ray morphology. We will assume in the later discussion that the
observed morphology corresponds to the actual distribution of soft X-ray
emitting gas.

\subsection{The lobes}

Beyond the plumes we observe large, lower surface brightness extended 
oval regions (which we will refer to as the E and W `lobes') which lie
along a position angle of 100-110 degrees. This larger scale emission
extends across 25 kpc from the end of one lobe to that of the other, and 
each oval lobe is 8 kpc along its major axis. The east lobe has a major axis
diameter of 23 arcseconds (8 kpc) along PA 90 degrees, and a minor axis
diameter of 15 arcseconds (5 kpc). The west lobe is elongated NE to SW
and may consist of two separate components; it is 10 x 22 arcseconds in
size. We have estimated nominal centers of the east and west lobes at
11 and 8 kpc from the nucleus.

The smoothed image makes it appear that the lobes are edge brightened, 
although this is only marginally evident in the raw data and one might
worry that it is an artifact of the smoothing process.  Extracting
counts from raw data in the lobe centers and the lobe rims shows a
difference of 4 sigma, with 6 total counts in the 0.2-2 keV range for a
pair of 5 arcsecond diameter circles in the E and W holes (lobe centers)
respectively,
compared to 26 total counts in a bright rim region of equal size. We
conclude that the lobes are indeed edge-brightened, although an exposure
three times as deep would provide a firm confirmation.

The lobes show marginal evidence for some very soft photons below 0.25
keV, but the spectral calibration of ACIS is still unreliable in that
energy band. The spectrum is consistent with no absorption in
excess of the foreground 
value from Stark et al.; formally, the best fits of the east lobe data to 
a two
temperature Raymond-Smith model are with a lower temperature of
$0.3\pm0.1$ keV and a higher temperature of  $1.0\pm0.2$ keV, while
the west lobe is slightly softer with values of 0.2 and 0.7 keV.

There are 364 total net counts in the lobes. Their total luminosity is
$1.3\times10^{40}$ erg s$^{-1}$, a result which is insensitive to the 
assumed
spectral fit parameters (but see below). We fixed the absorption at the 
galactic value
of 4$\times10^{20}$ cm$^{-2}$; we were not able to obtain an acceptable
fit with a single Raymond-Smith plasma for either lobe, but the sum of
two temperatures was adequate (although of no physical significance with
this small number of counts).

We note that ACIS-S is not very sensitive to diffuse hard emission.
The hard background in our data gives a weak upper limit of
about $1\times 10^{40}$ erg s$^{-1}$ for the 2-10 keV luminosity in the 
lobes,
comparable to the observed soft luminosity. Therefore, although we
have demonstrated the presence of low temperature X-ray gas in the lobes,
we cannot rule out the presence of hotter temperature gas.
Nevertheless, we see no evidence for the $1\times 10^{41}$ erg s$^{-1}$
of emission reported by Iwasawa et al (2001) from BeppoSAX data. This
emission was detected in a 3 arcminute circle and is not inconsistent
with our upper limit on this scale, but we can rule out its association
with the inner regions of Arp 220; possibly they underestimated
the contribution from Heckman's group (see below). The spectra
of the lobes and the plumes are compared in Fig. 7.

\subsection{The background cluster}

Fig. 8 shows a larger scale X-ray image which includes the region of
Heckman et al.'s (1996) z=0.09 group 1RXH J153456.1+232822. The X-ray
observations of this object and others in the field will be discussed in
detail in Paper III (McDowell et al., in preparation). We briefly note here
that we confirm the X-ray source as a diffuse object associated with the
optical group and detect several embedded point sources.
Galaxies Ohyama A and C (Ohyama et al. 1999, Grogin \& Geller 2000),
which are are also 2MASS sources (2MASS, 1999),  are both bright
X-ray sources, with Ohyama A showing a soft extended plume, while the
third cataloged galaxy, Ohyama B, is not detected.  Two further bright
point sources CXOU 153452.9+232833 and CXOU 153451.9+232828 are present
within the cluster but have no counterparts on the POSS2 images.

\section{Correspondence with other wavebands}

\subsection{Optical line emission}

We obtained new integral field spectroscopy of the Arp 220 system
in H$\alpha$ and NII as part of a program to study ionized gas in
ULIRGs. Arribas, Colina \& Clements (2001) presented observations
of the nuclear region of the galaxy. The present observations
were taken on 2000 May 8 and 9 on the 4.2m William Herschel Telescope
using INTEGRAL (Arribas et al. 1998) and WYFFOS (Bingham et al. 1994) and
consist of a mosaic of three pointings with a 30 arcsecond field-of-view
fiber bundle.

As in other star forming galaxies studied by Chandra (e.g. Strickland et 
al 2000), there is 
a strong correlation of X-ray morphology with optical line emission.
Heckman, Armus \& Miley (1987) discovered the 20 kpc `double-bubble' 
H$\alpha$
structure around Arp 220, as well as the inner H$\alpha$ SE plume.

The H$\alpha$ and soft X-ray morphology agree well in overall form and 
location - faint
lobes, stronger plumes and the bright nuclear region - but there
are differences in detail. Fig. 9 shows contours of H$\alpha$ emission 
superimposed on
the soft X-ray image.

The peak of the H$\alpha$ emission coincides to within the 1"
registration error with the soft X-ray peak (X-3).  The holes in the
lobes agree to within a couple of arcseconds, and the eastern  plume is
brighter than its western counterparts in both H$\alpha$ and X-rays.
However, the brightest part of the line emission in the western lobe is
shifted about 8 arcseconds south of the western X-ray lobe, and the
eastern lobe is much more prominent in  X-rays than it is in H$\alpha$. 
The peak of the SE plume is 2 arcseconds north of the plume's soft X-ray
peak.

For a typical electron density of $100 \mbox{cm}^{-3}$ and temperature
of $10^4$ K we derive the mass of H$\alpha$-emitting gas to be  $2\times
10^6 M_\odot$. The velocity in the plumes derived from the H$\alpha$ 
observations (Colina et al 2003, in prep.) is about 200 km/s on average
(with gradients up to 600 km/s in the lobes). Interpreting this as an
outflow leads to a kinetic energy of $8\times 10^{53}$ erg s$^{-1}$.

\subsection{Overall picture of Arp 220}

Arp 220 is a complex system with tidal tails, dust lanes, and multiple
nuclei. About 150 clusters and nine satellite galaxies can be seen on a
recent HST I-band image (Paper III), reinforcing the
impression that the Arp 220 system is massive compared to the
Milky Way.
The hard source X-1 is clearly associated with the galaxy's central
regions, and
is probably coincident with the western nucleus Arp 220B (Paper I),
although uncertainties in registration mean we cannot rule out an
identification with the eastern nucleus.  The soft emission and the X-3
peak (which also has a soft spectrum) coincide with the emission line
peak rather than with the optical continuum peak. As other authors have
discussed, the location of the optical peak is likely determined by
gaps in the absorption and may not correspond to a physical object. None
of the known optical and infrared clusters in this system (Shaya et al. 
1994, Scoville
et al. 1998) are detected in X-rays. X-2 is in a star cloud just to the
south of the western end of the central dust lane; there is a probable
cluster at 1.2 arcsecond from its estimated location, not close enough to
propose as an identification.

The optical dust lane coincides with the separation
between the nuclear emission
and the SE plume, indicating that this separation is probably due to 
absorption.
Interestingly, the plumes' outer boundaries, while
not sharp, coincide with a drop in the optical isophotes and plausibly
represent the escape of the gas into a less dense halo interstellar 
medium.
The axis of the plumes is perpendicular to the dust lane of Joy et al.
(1986) and the CO disk, implying that they may be collimated in the
inner region. One might expect the faintness of the NW plume to reflect
its orientation deeper into the galaxy, but the X-ray spectrum does not
allow us to constrain the possibility of larger absorption in the NW.
However, a second dust lane visible in the HST data 
bounds the brightest part of the NW plume on its northern side,
implying that there may be more X-ray emission hidden behind it.

\section{Discussion}

\subsection{The Superwind and the Plumes}

Heckman et al. (1996) proposed that the extended H-alpha and soft X-ray
emission in Arp 220 originates in a `superwind' driven by starbursts 
which have been triggered by the merger. An alternate possibility, to 
be discussed in more detail by Hearn \& Lamb (2003), is that the
galaxy collision and merger dynamics are directly responsible for the 
extended structural morphology and hot gas. We consider each of these 
possibilities and conclude that both processes may play important 
roles in the Arp 220 system.

If Arp 220 is the product of the merger of two gas-rich disk galaxies, 
as is commonly thought to be the case because of the apparent presence
of  two, very close galactic nuclei and the overall disturbed
morphology,  then  the physical processes that take place in this
collision lead not only to conditions ripe for a central starburst, but
also large-scale shocking of  the interstellar gas over significant
parts of the galactic disks. That  is,  in such a system, it is expected
that a starburst-generated superwind  will  have to punch through a
complicated region that lacks three dimensional  symmetry, and consists
of sectors of rapidly outflowing and inflowing shock-heated gas, 
coupled to sectors of flowing neutral gas that did not experience a 
direct impact. Thus the dominant flow direction of the superwind is 
likely  to be constrained by the evolving geometry of the colliding
system.  The plumes will not expand into the regions  of  high pressure
and high density, that comprise the hot expanding shocked  gas  and the
cool expanding gas, respectively. The three dimensional  morphology of 
the overall system is very dependent on the collision parameters, and
the time since first impact. Thus the environment of a superwind in a
merging  system  is possibly different to that in a system like M82,
that appears to be  experiencing a central massive starburst in a
previously relatively  undisturbed disk galaxy.

Strickland and Stevens (2000) have used a combined Eulerian/Lagrangian 
hydrodynamics code to perform 2-D modeling of the superwind in M82, 
assuming  cylindrical and reflection symmetry. As discussed by them,
construction  of  a model to match the observed wind proved to be
elusive, and will  possibly  require the inclusion of processes such as
inflow of cold material or  magnetic fields, to produce the observed
confinement at the base of the  wind in the starburst region. They
conclude from their modeling that the  bulk of the outflowing gas is at
high temperatures of around $10^{7.5}$K and low density, and would,
therefore, be difficult to observe because of  low emissivity.
Interesting new information on the nature of  starburst-driven  winds
and the interpretation of their observed X-ray fluxes is provided  in 
Strickland et al. (2000). This tends to support the supposition that the
temperature of the outflowing gas is not easily determined from X-ray 
data. They find that high-resolution Chandra observations of the 
kiloparsec-scale wind in NGC 253 demonstrate a strong correspondence 
between  the X-ray and H-alpha observations of the outflow cone, with 
approximately  equal amounts of energy being emitted from the hot gas in
optical lines  and  in X-rays, and a total energy in both of the order
of $10^{41}$ erg s$^{-1}$.  This is  about one percent of the mechanical 
energy estimated to be injected by  the  starburst. Strickland et al. (2000)
also determined that the bulk of the  X-ray  flux originates in a
limb-brightened structure, with at most 20\% of the  flux coming from
the body of the fluid. This provides the first direct  evidence that
this radiation is generated at the interface between  the outflowing gas
and the ambient denser medium, rather than in the bulk  of the flowing
gas, whose temperature is, therefore, not well established.

Comparing our Chandra observations of Arp 220 with those of NGC 253, we 
see  some similarities, despite the large difference in scale. As noted
previously,  the extended X-ray lobes in Arp 220 extend a distance of
10-15 kpc from the nucleus on the plane of the sky, and 
these also appear to be limb-brightened. The  observed ``edges'' show 
reasonable overlap with the arcs of H-alpha emission in  these  regions, 
one each on the east and west sides (see Fig. 9). The lobes  also  
display apparent central "holes" in the emission. In fact, the counts in  
the  lobe regions are sufficiently low that our results are consistent 
with  all  of this soft
X-ray flux originating within edge-brightened structures.  However, 
although both sets of structures (plumes and lobes) appear to be produced 
by outflowing gas colliding with an external, denser medium, this may be 
an incorrect model for the lobes of Arp 220, because simulations of 
colliding
galaxies  indicate  that the collision itself may have flung a dense
ring-like structure of gas out on each side of the nucleus. For this
system, we suggest that at these vast distances from the nucleus, it
is the tidal interaction itself that is propelling the gas outwards;
that there is shock heating within, and along the edges of the locally
overdense structures, as demonstrated in  previous  studies of colliding
galaxies (see Gerber et al, 1996, Lamb et al, 1998, and Hearn \& Lamb, 
2001); and  that whether the source of the outflow energy is overpressure 
from a central  starburst
or tidally channeled galactic gravitational energy, the net  effect  is
a strong correlation between the X-ray emission and H-alpha emission 
from  gas that attains a temperature of around $10^6$K in shocks at an 
interface.

The inner two plumes of hot gas observed in Arp 220 are likely to be due
to a superwind that is driven by a central massive starburst, because
such a wind would be a natural outcome (although similar features can be
produced by flowing shocked regions of gas in slightly off-center galaxy
collisions). In this first picture, the central regions have become
sufficiently dense that a large fraction of the gas has cooled to form
molecular gas and a super starburst has occurred. This latter has then
produced a superwind which has interacted with, and been somewhat
chanelled by the structures formed by the collision itself.  

The extent of these plumes is large, spanning 11 kpc across
the nuclear region. However, in contrast to the outflow cone in NGC 253,
the Arp 220 plumes are very bright in soft X-rays, and no
limb-brightening is noticable in our data.
 The ambient medium into which these plumes are driving may be
considerably different to that in NGC 253, consisting of entangled
regions of shock-heated flowing gas and cooler, non-shocked gas, as
discussed above. The energy flux coming from the central
starburst in Arp 220 is larger than that from NGC 253 by
a factor of 70, and this may also play a role in producing the different
X-ray properties of the two objects. As noted in Section 3.3, the
Chandra X-ray data for the Arp 220 plumes is consistent with a  thermal
plasma with temperatures ranging from 0.25 to 1 keV. This temperature
range may result from the interaction of gas ejected in hot stellar
winds and SN with regions of collisionally shock-heated gas, and with
cooler flowing gas. That is, the entrained material may be of very
varied initial temperature.

\subsection{Merger Simulations}

The overall context within which we wish to discuss our Chandra X-ray 
data for Arp 220 is that of a colliding pair of co-rotating, comparable 
mass, gas-rich galaxies which have not yet fully merged, but whose 
nuclei are now very close, and lie within a central disk of material 
that has been formed from remnants of the original two gaseous disks.
Other disk remnants have been flung to large distances from the joint 
nuclear region. Relevant numerical modeling (Lamb et al 2003)  indicates
that the collision was likely almost face-on, and had an  impact
parameter of roughly 70\% of the optical disk radius. Such a collision
will leave some parts of the gas disks, those furthest  from the impact
point, unshocked by the direct collision, and this  material is likely
to retain its original temperature, whereas that  gas involved in the
direct collision between the two disks will have  been heated to over
$10^6$ K. All regions experience an initial inflow  towards the impact
point (that point in each disk where the center  of mass of the other
galaxy passes), followed by outflow, and then  further inflow (see
Gerber, Lamb, \& Balsara, 1996). The subsequent,  gravitationally-driven
flows lead to relative gas velocities that  are much larger in magnitude
than the original impact velocity,  with relative velocities of
approximately 900 km/s obtained in  some regions of the disk for the Arp
220 model fit, assuming that  the Arp 220 system has a mass similar to
that of the Milky Way.  The rate of inflow and outflow in the disk
material scales with  the initial distance of the material from the
impact point, being  smaller for the more distant stars and gas. This,
together with  the asymmetries introduced by an off-center collision,
lead to  an evolving, intricate structure in the gas. As has been noted
by  many previous authors, such collisions can lead to the accumulation 
of large masses of gas in the merging nuclear region (for example,  see
Mihos and Hernquist, 1996). We note that this can only happen  on a
timescale of relevance to a system such as Arp 220 if the  initial
impact velocity is relatively low, less than a couple  of a hundred km
s$^{-1}$ for galaxies with a mass of several  $10^{11}M_\odot$, because
larger impact velocities lead to a  much longer merger time for the
system as a whole.

The particular detailed numerical simulations of a pair of colliding
gas-rich disk galaxies which was found to provide a suitable fit to the
large-scale structure of Arp 220 was taken from a series of simulations
in which mass ratio, impact velocity, impact parameter, relative disk
spin direction, and relative tilt-angle of the two galaxies have been
explored (Lamb et al., 2003). The  3-D simulations include a
representation of the gas, stars, and dark  matter, which all interact
gravitationally. The gas hydrodynamics is  followed using the method of
Smooth Particle Hydrdynamics (SPH), and the  gravational potential is
calculated on a grid for the full particle set.  A description of the
details of this n-body/SPH methodology, together with the galaxy
starting models,  is given in  Gerber, Lamb \& Balsara, 1996. We note
that in these simulations, the  gas is considered to be isothermal. That
is, there is no explicit heating and cooling in the simulation
calculations. This is a useful approximation if one wishes to follow   
the overall dynamics of the gas and the build-up of dense regions, but 
we are not able to give a detailed map of the eventual temperature
structure in the system. Rather, we can give a general description of   
the eventual location of that gas which has passed through shocks during
the collision, and that gas which has not. An example of the      latter
is the gas that is in the outer edges of the disks that do not   
overlap at impact because of the off-center nature of the collision.
This material is rapidly propelled outwards in the plane of the disks, 
after the brief overall contraction of the two galaxies. 

A detailed exploration of the temperature distribution in the  system as
it evolves must await further modeling with appropriate  physics
included. However, it is of interest here to explore the  range in
possible post-shock temperatures obtained in the gas due  to the
collision. Assuming that the strong shocks that we have here  are
adiabatic, we can use the standard analytical expressions that  relate
the post-shock temperature of the gas $T_{2}$ to the relative  velocity
$v$ of the colliding gas streams, and the adiabatic index  $\gamma$ and
mean molecular weight $\mu$ of the gas (see Landau \&  Lifshitz 1987,
Chapter~9). For a $\gamma = 5/3$ gas,  $k_{\rm B} T_{2} = \mu \times
{v^2}/3$. In the present situation,  $\mu = 0.6 m_{\rm H}$, where
$m_{\rm H}$ is the mass of the hydrogen  atom. Thus for the maximum
relative velocity found in our simulation  of 900 $\mbox{km~s}^{-1}$, we
estimate the post-shock temperature  is $\sim 2 \times 10^7$~K. However,
we expect the bulk of the  shocked gas to be at a few million degrees
because the relative  velocities of the internally-generated gas streams
range downwards  from the maximum. For a relative velocity of $200
\mbox{km~s}^{-1}$,  the post-shock temperature is just $10^6$~K.

In Fig. 10, we present a view of the gas distribution in our best-fit
3-D  model from a viewing angle appropriate to our line-of-sight to Arp
220.   The gas density is represented by intensity (color in the
electronic   edition), as indicated in the figure caption.  From an
inspection of the  morphology of the simulation,  we identify regions
that correspond to the  lobes and plumes observed in Arp 220, as shown
in the figure. We also  indicate the general regions in which mostly
unshocked gas resides at  this time in the  evolution of the system. We
connect these latter  regions to those in Arp 220 that show HI emission,
such as those observed  by Hibbard, Vacca \& Yun (2000).

The choice of model is constrained in both computational simulation 
epoch and in viewing angle.  The distinct changes in the morphology  of
the simulated galaxies over relatively short time scales due to  the
collision dynamics produces the observed features only over a  select
range of times after the collision.  There is some flexibility  in the
viewing angle chosen, but a very small fraction of the $4\pi$ 
steradians is available to produce a match.  Projections of the  complex
three-dimensional structure of this collisional system onto  the plane
of the sky necessarily produces a wide range of distinct  features that
depend upon the direction from which the system is  viewed.  This
restriction in viewing angle is a considerable strength  of the model in
that it places strong constraints on the predicted  integrated
line-of-sight velocities across the structure; these  velocities can be
compared to kinematic observations of Arp 220  when they are available.

We note that the simulations include the effects of gravity and  gas
dynamics on the preexisting stellar and gas distributions, but  do not
include the feed-back effects of star formation or a   superwind on the
gas dynamics. Nevertheless, the model shows that   structures similar to
those observed in Arp 220 can arise purely  from the  merger dynamics.
We have demonstrated in previous studies  (see Hearn \& Lamb, 2001) that
the dense regions that form in the  gas due to a collision are prime
sites of episodic, large-scale  star formation events in colliding
galaxies. Thus, we expect that  star formation has taken place in the
dense features that formed  in Arp 220. Where young stellar populations
have been formed,  they will dominate the blue light, and will likely
dominate the  V-band also.

The maxima in the gas densities obtained in this simulation at  this
time in the outwardly propagating rings that form the outer  edges of
the "lobes" are approximately $2.2$ and  $2.4 \times 10^6
\mbox{$M_\odot$ kpc$^{-3}$}$ for the two  respectively, again using
Milky Way Galaxy scaling. The bulk  of the material in these structures
is at lower density, and  $10^6\mbox{$M_\odot$ kpc$^{-3}$}$ is a good
average value.  This is to be compared to a maximum in the inner regions
of  the model system of $3.8 \times 10^7 \mbox{$M_\odot$ kpc$^{-3}$}$ 
which occurs at a distance of about 3 kpc from the gravitational  center
at this epoch. All densities calculated in these  simulations scale with
the mass of the system.

The numerical models also provide a wealth of information on the
velocity structure. However, the absolute values of the velocities and
the time parameter in the simulations scale with the system mass and 
radius. In general, the computational units are chosen such that the 
quantity $G M T^2 / L^3$ is a dimensionless scalar, where G is the 
gravitational constant, and T and L are the computational time and
length  units, respectively, and in these simulations, $M T^2 / L^3$ is
set equal  to 1. Thus $T = L \sqrt{L/M}$, indicating that the
gravitationally driven  processes occur most rapidly in compact, massive
galaxies. We do not have  direct knowledge of the original radii of the
two disk galaxies involved  in the Arp 220 merger, nor do we yet have a
very precise mass estimate  for this system, therefore we will discuss
the relative velocities  generated in the model, and provide magnitudes
based upon the mass and  radius of the Milky Way galaxy for reference.
We see from above that the  scaling of the time unit is not very
sensitive to the mass; we also note  that the value of the length unit
can be constrained by detailed  comparisons of the models and
observations of a particular system. In  this case, the best fit model
shown in Fig. 10 corresponds to a time  $1.5\times 10^8$ yr after the
closest approach of the two nuclei, making  Arp 220 a very young,
incomplete merger. 

These remarks should not be construed as implying  that the full time
interval of approximately  $1.5\times 10^8$ years is available for a
superwind to propagate out  through the galaxy from the central nuclear
region seen at this epoch in the model. Only after the joint nucleus is
formed will enough stellar  material be colocated over a long enough
time to allow the enhanced star formation and merged superwind
associated with a nuclear super-starburst.  Although dense, presumably
star-forming regions form in the gas in each disk very soon after the 
collision, their location in the disk changes over time, and none of
these regions reform into the two galactic nuclei until well after the
initial closest approach of the  two nuclei. It takes an even longer
time for the material from both disks to settle into one, joint central
region, as observed in the Arp 220  system. The time for this central
stellar turn-on is well defined in the sense that the central density
remains low until this time and grows rapidly thereafter. In the chosen
model, with the Milky Way scaling applied, the combined central region
has been in existence for approximately  $1.5\times 10^7$ years, a
factor of ten shorter than the time since the  first closest approach.
This then provides the timescale for the outward propagation of a
superwind from a starburst in this central location. The  central region
will have been fed by a continual stream of infalling  material during
this $1.5\times 10^7$ year period, which might be  expected to help fuel
an increasingly larger central starburst. This  result is consistent
with those of Mihos and Hernquist (1996) in their  investigations of the
timescale for central merger, infall of gas, and  the potential
production of a central starburst, in systems with somewhat different
collision parameters than those presented here.

The regions in the model corresponding to the `plume regions' in Arp 220
show an infalling of gas towards the nucleus of about 400 km s$^{-1}$, 
rather than the outflow of 225 km s$^{-1}$ derived by Heckman et al 
(1996) for a superwind in this system. We note that both of these values
 would be reduced by projection effects as viewed on the sky and, at the
 approximate viewing angle given by our model fit, become approximately 
300 km s$^{-1}$ and 170 km s$^{-1}$, respectively. Neither of these 
numbers should be given great weight because both are based on simple 
assumptions for the flow mechanism and the properties of the system. 
However, the interaction of an infalling collisionally shocked gas
stream  with a superwind would provide an interesting dynamical
situation in the  central regions of the galaxy and could lead to a
slowed wind or to  turbulent mixing, for example. In such a case the
usual boundary  conditions used in superwind calculations, such as those
of  Strickland \& Stevens (2000) would need to be modified, and a  3-D
simulation would likely be needed.

The outer, dense edges of the `lobe regions' in the model are continuing
to expand outwards and are experiencing a general rotation that is a  
remnant of the original rotation of the two disks. The rotational
velocity is larger than that of the corresponding Keplerian velocity at
that radius because a collision leads to an increase in the azimuthal
velocity of the expanding ring (see Gerber 1993, and Gerber et al.
1996). The combination of the azimuthal and radial motions for these   
rings leads to a declining line-of-sight velocity along each of the lobe
rims ranging from approximately 185 to 25 km s$^{-1}$, and -175 to -20
km s$^{-1}$, in the frame of the galaxy, for the two respectively.
Insufficient observational data for these regions in Arp 220 currently  
exists to check these predictions against observations, but the   
detection of consistent rotation in the two lobe exteriors would provide
support for our model.

Fig. 11 displays the distribution of the disk star particles
corresponding to the same time step and viewing angle as shown for the
gas distribution in Fig. 10. The particles representing the original   
disk stars are shown in black (yellow in the electronic version). We    
note that the distribution of the stars has an overall similarity to
that of the gas, as both are driven primarily by the evolving
gravitational potential. However, the density features in the gas are
sharper than those in the stellar distribution because of the
collisional nature of the gas, and the size of the expanding gaseous   
envelope is slightly smaller than the expanding remains of the stellar
disk. In a real galaxy, we expect that the old original disk stars would
contribute heavily in the near-IR.

If we can assume that the original gas disks which formed Arp 220 did
not extend far beyond their respective stellar disks, which is not a
good assumption for all disk galaxies (see Briggs et al. 1980, Martin
1998), then we predict that  the gas and the old, intermediate and
low-mass stars should be found to roughly co-exist in space on large
scales. That is, we expect the near  IR flux from these stars in Arp 220
to extend as far as the X-ray and H-alpha images, or a little further.
The collisional nature of the gas  produces high density features not
found in the stellar mass   distribution, so the detailed structure is
not the same in both, and the old stellar distribution in such systems
can appear relatively smooth (see Stanford \& Bushouse 1991) in
comparison to that of the gas, and even more so in comparison to that of
the new star-forming regions.

If either of the original disk galaxies had an extensive HI disk beyond
the stellar disk, much of this material would now contribute to
distended HI features on either side of the newly forming tight, central
disk observed in the system. However, some fraction of the original HI
gas, that residing on the near sides of the two disks during the
collision, would have been collisionally shocked and heated to high
temperatures. Galaxy collisions provide a means of heating  large
amounts of HI gas, as well as delivering vast amounts of material to the
galaxy center, or into other dense regions, such as the extensive ring
structures and tidal arms observed in some colliding systems (see Lamb,
Hearn \& Gao 1998, and Hearn et al. 2001), that can cool efficiently and
subsequently form extranuclear superstarclusters. Indeed, Arp 220 does
display such starclusters away from the nucleus, and the location and
age of these can help to pin down the correct   numerical models for
this system, and even help to put constraints on the masses of the
colliding galaxies.

\subsection{The Lobes as Superwind or Merger Products}

The lobes are morphologically distinct from the plume regions; their   
lower surface brightness and different position angle indicate that   
either the lobes and plumes are dynamically distinct, or that the
outflowing material undergoes a transition at the lobe/plume boundary. 
In the latter case, we can interpret the lobes as the outflowing
material from the plumes which has encountered a rapid drop in density
at this radius in the galaxy's halo and created a low-density, hot
bubble of gas on either side of Arp 220. In principle, an unimpeded
superwind of velocity $1000v_3$ km s$^{-1}$ could create a bubble of 
size $10r_{10}$ kpc in $1\times 10^7 (r_{10}/v_3)$ yr where $r_{10}$
and $v_3$ are scaling parameters. This would be feasible, given the time
 scaling of our merger simulation, if we assume that we do not observe
the lobes in projection, which would imply a larger true linear  
dimension, and that the central starburst occurred immediately after
the combined stellar and gaseous nucleus formed. The numerical
simulations indicate that the material now found in the central regions
of Arp 220 was initially flung to a considerable distance from its   
current location, and has subsequently fallen back inwards in a steady
stream that continues to this epoch. It appears unlikely that a strong,
unimpeded superwind has existed for the last $10^7$ yr. Further, if the
true superwind velocity is low, as suggested by Arribas et al (2002),
the timescale for lobe formation would be considerably longer than the
maximum of $10^7$ years available in the collision and merger model
described here. If the true mass of the Arp 220 system is larger than
that of the Milky Way, the time available for the superwind to
propagate outwards is decreased.

In this section, we consider the alternative possibility that the     
connection between the lobes and the plumes is an artifact of projection
 and that the lobes arise purely from merger dynamics and collisional  
shock heating of the gas. As we mentioned earlier, the particular
morphological form that appears to provide a good fit to the observed 
large-scale structure of Arp 220 can be obtained for off-center,  
face-on, co-rotating disk collisions if the impact velocity is less
than a couple of hundred km s$^{-1}$. If the impact velocity is larger,
for example 300 km s$^{-1}$ or more, lobes form and then disperse long
before the nuclear region has had time to recondense, and we would not  
then expect either a starburst or its accompanying superwind to coexist
with the collisionally induced lobes. The expanding, rotating,   
non-planar rings obtained in our chosen model, those that we identify
with the outer, limb brightened portions of the lobes of Arp 220, do
not appear ring-like from the chosen viewing angle. The gas contained
within them was once a part of the two disks and has experienced both a
collision with the gas from the other galaxy and, internally, with
infalling, outer gas from the same disk. This latter process often leads
to large-scale clumping of the gas, and a highly asymmetric density
distribution around the ring (see Gerber, Lamb, \& Balsara 1992), as
well as very high relative velocities. In this picture, our
line-of-sight to the centers of the lobes, as projected on the plane of
the sky, passes through mostly very low density gas because the
outwardly propagating ring structures have excavated these regions. The
details of this proposed origin of the Arp 220 lobes can be further
explored and checked against observations once detailed dynamical
information for the lobes is available, and once a better mass estimate
for the system is available from near-IR photometric imaging.

\section{Conclusion}

We have demonstrated the presence of distinct regimes in the X-ray
emission in Arp 220: a nuclear region where a mixture of sources
contribute (discussed in Paper I); a plume region of end-to-end
dimension 11 kpc, with X-ray temperatures from 0.2 to several keV, that
we suggest is likely associated with a superwind from the starburst at
the center of the merger; two cooler diffuse lobe regions outside the
optical galaxy  with a temperature of 0.2-1.0 keV, that stretch to 10-15
kpc on either side of the nuclear region. We suggest the lobes are a
product of the  collision itself, when a fraction of the gas gains a
sufficiently higher than average share of the collision kinetic energy 
to radiate in the observed X-ray range. This picture helps to resolve
several of the puzzles about this system raised by Heckman et al (1996);
namely, the  'misalignment,' by 25-30 degrees, of the plumes and lobes,
by  attributing each to different structures rather than to a continuous
 structure, their apparent continous appearance on the sky being caused
by  superposition; the integrity of the outer lobes, which we interpret
as  collisionally produced ring structures, rather than the outer edges
of a  wind-blown bubble which might be expected to have experienced
'blow-out'  by this time; and the collimation of the supposed wind-blown
plumes which  may have been influenced by the collisionally modified
phase and velocity  structure of the near-disk galactic environment. We
note that the disk  itself is likely to have only recently reformed
after the disk disruption  caused by the initial collision of the
galaxies.

Arp 220 is a prototype for the large number of rapidly star-forming 
infrared mergers at higher redshift. The complexity of its interstellar 
medium is a salutary warning of the difficulty of interpreting data with
 poorer linear resolution. If the Arp 220 X-ray lobes are due to a 
superwind, the amount of energy being pumped into the intergalactic 
medium is a significant fraction of the energy generated by the
starbust,  in contrast to that of the superwind of NGC 253  which has an
energy of just  a few percent of the output from the central starburst. 
If the lobes are  a dynamical collision remnant, they may provide a
source of infalling  material at later times, but can also be dispersed
into the intergalactic  medium if the collision parameters and galactic
environment are  appropriate (Hearn \& Lamb 2003).

\acknowledgments

We acknowledge use of the ADS, SIMBAD, NED, and CIAO.

Partial support for this work was provided by the National Aeronautics
and Space Administration through Chandra Award Number GO1-1166 issued by
the Chandra X-Ray Observatory Center, which is operated by the 
Smithsonian Astrophysical Observatory for and on behalf of NASA under 
contract NAS8-39073. S. Lamb and N. Hearn acknowledge support from DOE 
contract LLNL B209032 and the support of NPACI through an 
allocation of computer time to perform the numerical simulations 
presented in this paper, and to NSCA for computational support for 
visualization of the models. We thank the referee for helping
us clarify the presentation.



\begin{landscape}
\begin{table*}
\caption{Flux of Arp 220 X-ray Components}
{\small
\smallskip
\begin{tabular}{lllllllll}
\\
Object                       & Position              & PA$^a$ & Extent      & Net counts& $F_{14}^b$ & $L_{40}^c$ & $L_{40}^d$ & log L(H$\alpha$)$^e$\\
                             & (J2000)               &        & ($\arcsec$) &           &(obs)    & (obs.)  &  (corr) & (erg s$^{-1}$)\\
\hline\noalign{\smallskip}
Arp 220 Circumnuclear        & 15:34:57.14 +23:30:13.0&-  &  7"      &  250 & 1.6 & 1.1 & 6 & 39.8 \\ 
Arp 220 X-1 hard halo$^f$        & 15:34:57.28 +23:30:11.4&-  &  3"      &  113& 5.9 &  4.1 & 8? &)      \\ 
Arp 220 X-1 (nucleus)$^f$      & 15:34:57.21 +23:30:11.7&-  &  Unres.  &   66& 3.0 &  2.1 & 4? &) 39.9   \\ 
Arp 220 X-4$^f$              & 15:34:57.25 +23:30:11.5&-  &  Unres.  &   19& 1.0 &  0.7  &1.5 &)     \\ 
Arp 220 X-2$^f$                  & 15:34:56.94 +23:30:05.5&-   & Unres.  &   33&1.0&    0.6&  0.7 &$<$38.0 \\
Arp 220 X-3                  & 15:34:57.14 +23:30:13.1&-  &  2"      &  31& 0.2 &   0.1 & 0.6&39.7  \\ 
Arp 220 SE Plume             & 15:34:57.64 +23:30:06.0&135&  7" x 14"&  409&3.3 &   2.3 & $>$2.5 &40.2 \\ 
Arp 220 NW Plume             & 15:34:56.96 +23:30:17.4&331&  8" x 9"&   135&0.8 &   0.5 & $>$0.7  &39.6  \\
Arp 220 E Lobe               & 15:34:59.21 +23:30:02.6&111& 23" x 15"&  188& 0.9 &   0.6 & 0.7 &39.3  \\
Arp 220 W Lobe               & 15:34:55.58 +23:30:14.0&280?& 10" x 22"& 176&1.0 &  0.7 & 0.8 &39.6  \\
Arp 220 remainder            &                        &    &           & 274&1.5&   1.0 & 1.2 &40.0 \\
\hline
Arp 220 total                &                        &    &           & 1694& 20.2& 13.8 & $>$26  & 40.7   \\
\hline\noalign{\smallskip}
\hline
\hline\noalign{\smallskip}
\end{tabular}

\parbox{5.5in}{
$^a$ Position angle in degrees of the line joining the feature to the nucleus.\\
$^b$ Observed flux in the 0.3-10.0 keV band in units of $10^{-14} \mbox{erg cm$^{-2}$ s$^{-1}$}$.\\
$^c$ Luminosity corresponding to observed flux in units of $10^{40}$ erg s$^{-1}$ (uncorrected for absorption).\\
$^d$ Luminosity corrected for X-ray absorption, in the same units.\\
$^e$ Log of H$\alpha$ luminosity (Colina et al in prep.)\\
$^f$ These are the only components with significant flux detected above 2 keV.
Observed soft flux for these components totals $4\times10^{39}$ erg s$^{-1}$.
}
}
\end{table*}

\end{landscape}


%


\begin{figure}
\epsscale{1.0}
\plotone{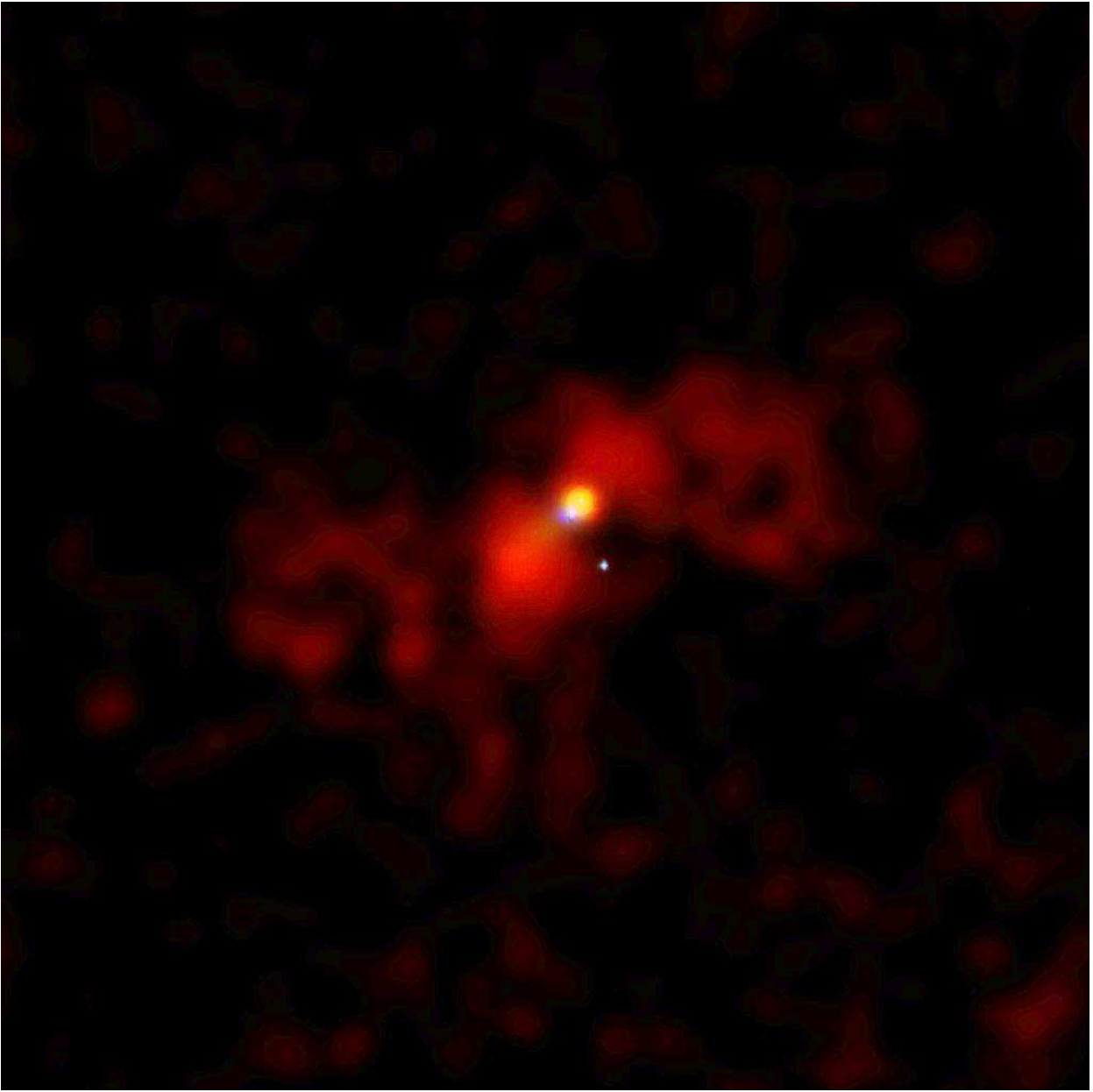}
\caption{\small Adaptively smoothed true X-ray color image of Arp 220.
Red represents the 0.2-1.0 keV band, green the 1.0-2.0 keV band, and blue
the 2.0-10.0 keV band. Field shown is 2.1' square.}
\end{figure}


\begin{figure}
\epsscale{1.0}
\plotone{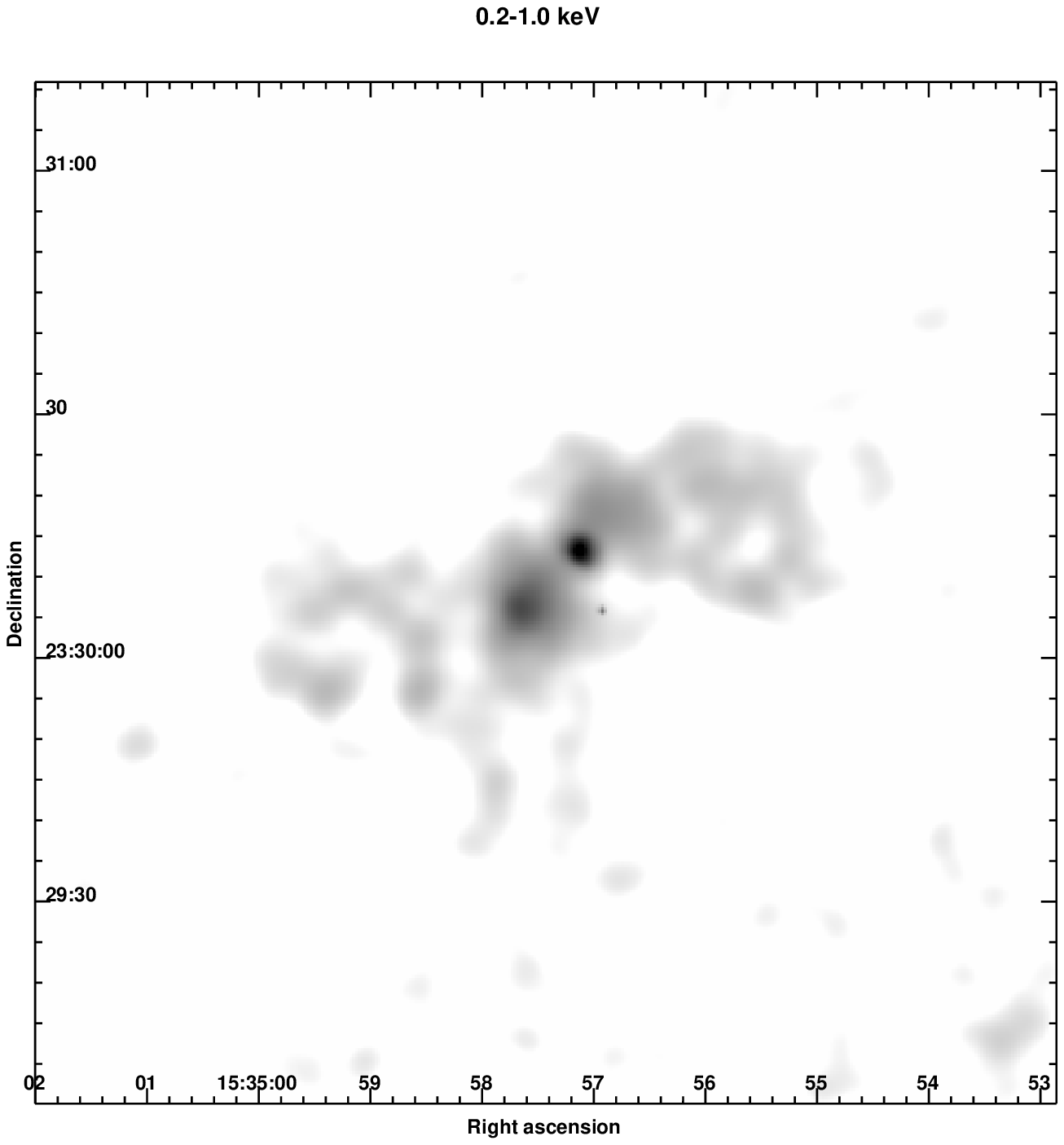}
\caption{\small Adaptively smoothed images of the Arp 220
region in three bands: (a) 0.2-1 keV}
\end{figure}

\begin{figure}
\epsscale{1.0}
\plotone{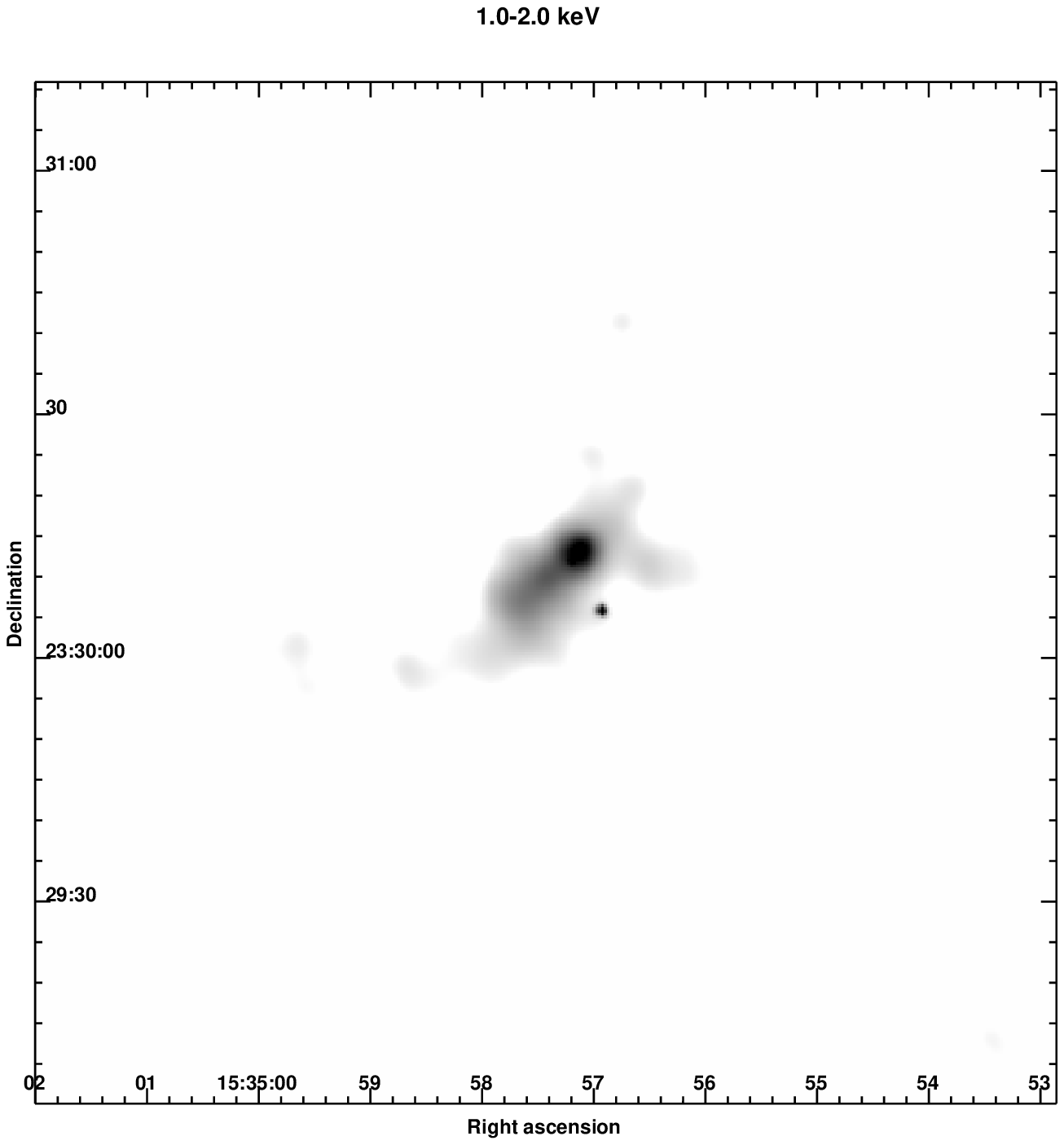}
\caption{\small Adaptively smoothed images of the Arp 220
region in three bands: (b) 1-2 keV.}
\end{figure}

\begin{figure}
\epsscale{0.5}
\plotone{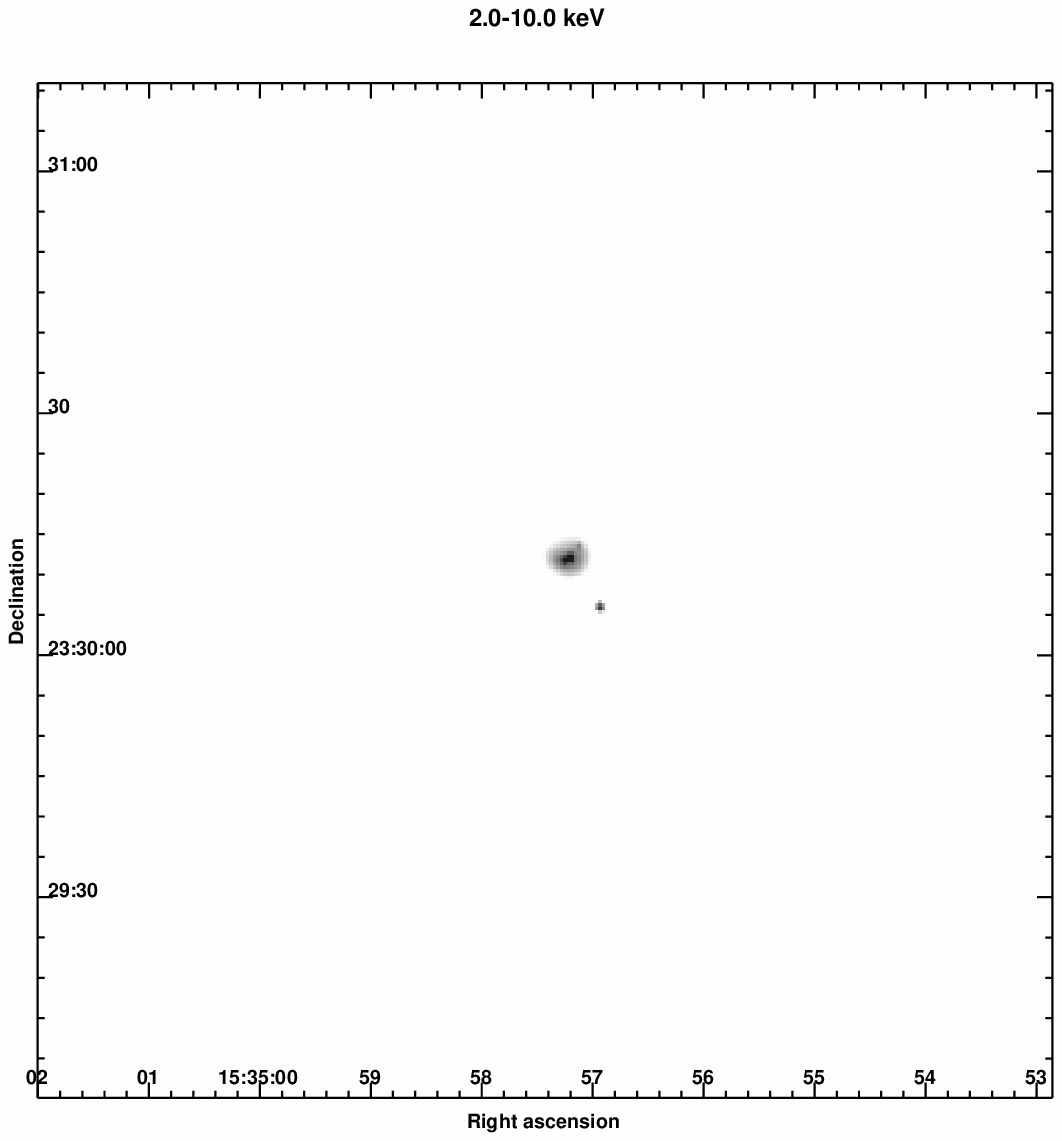}
\caption{\small Adaptively smoothed images of the Arp 220
region in three bands: (c) 2-10 keV.}
\end{figure}

\begin{figure}
\epsscale{1.0}
\plotone{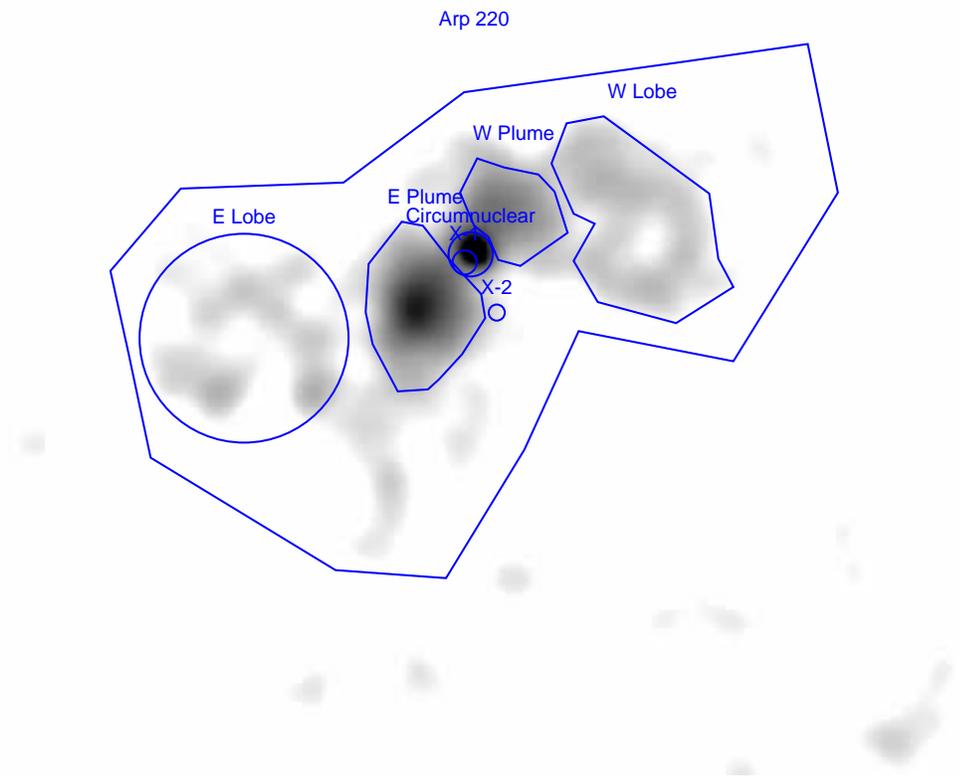}
\caption{\small Extraction regions used for spectral
analysis, overlaid on smoothed soft band image.}
\end{figure}

\begin{figure}
\epsscale{0.4}
\plotone{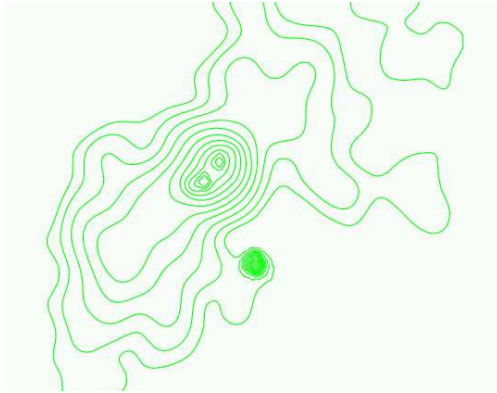}
\caption{\small Reconstructed image of central part of Arp 220,
showing plumes and circumnuclear region
in the 0.2-10.0 keV band. In each of the three energy bands, 
the three point sources X-1, X-2 and X-3 were subtracted,
the remaining emission was adaptively smoothed, and then PSFs
for the three sources were added back in.
The southeast `nucleus' is X-1 and the
northwest `nucleus' is the soft peak X-3; X-2 is the strong peak
to lower right of image.
Intensity contours are logarthmically spaced 0.2 dex apart.
}
\end{figure}

\begin{figure}
\epsscale{0.5}
\plotone{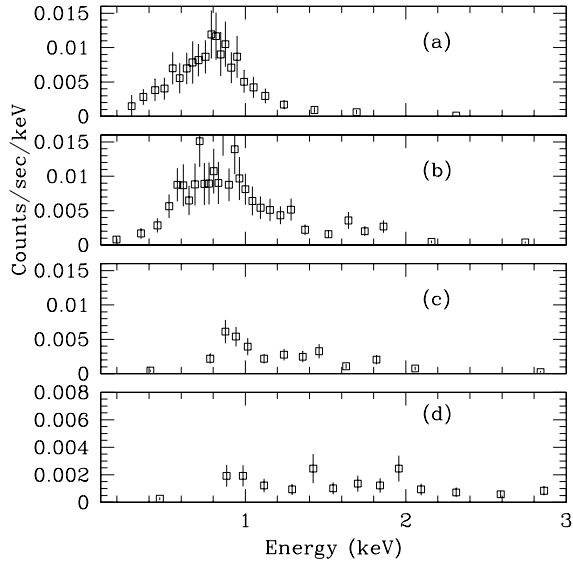}
\caption{\small Background-subtracted pulse height (PI) spectra
for (a) the lobes, (b) the plumes, (c) the circumnuclear region and source X-3,
and (d) the X-1 region. }
\end{figure}


\begin{figure}
\epsscale{0.6}
\plotone{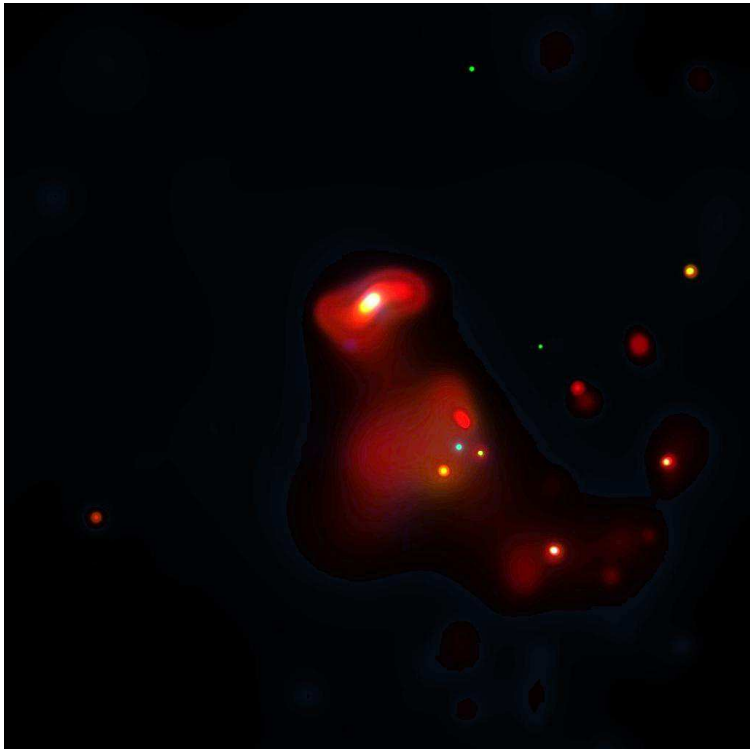}
\caption{\small The Arp 220 field (0.2-10.0 keV), adaptively smoothed
to bring out emission from the cluster.}
\end{figure}

\begin{figure}
\epsscale{0.6}
\plotone{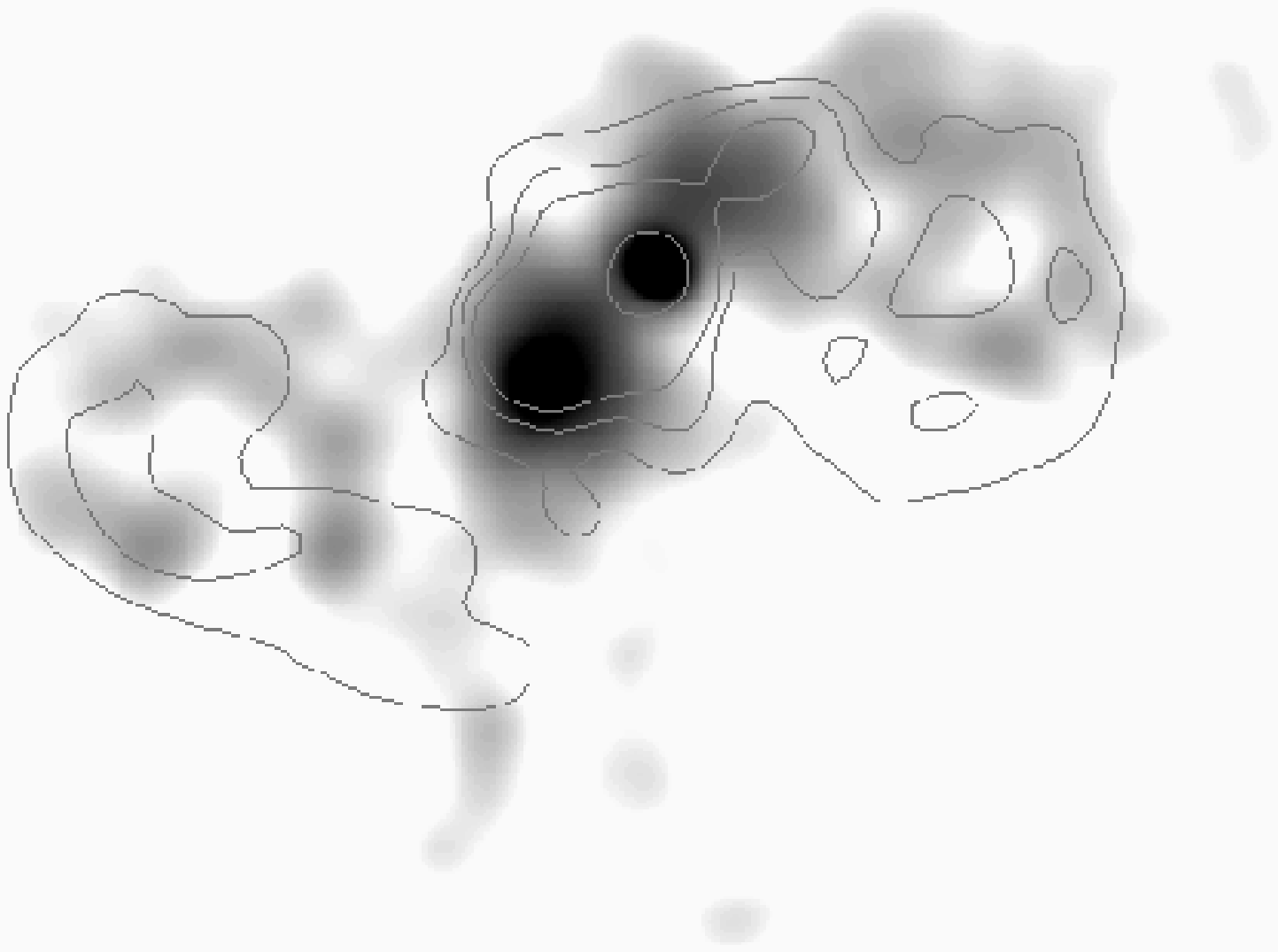}
\plotone{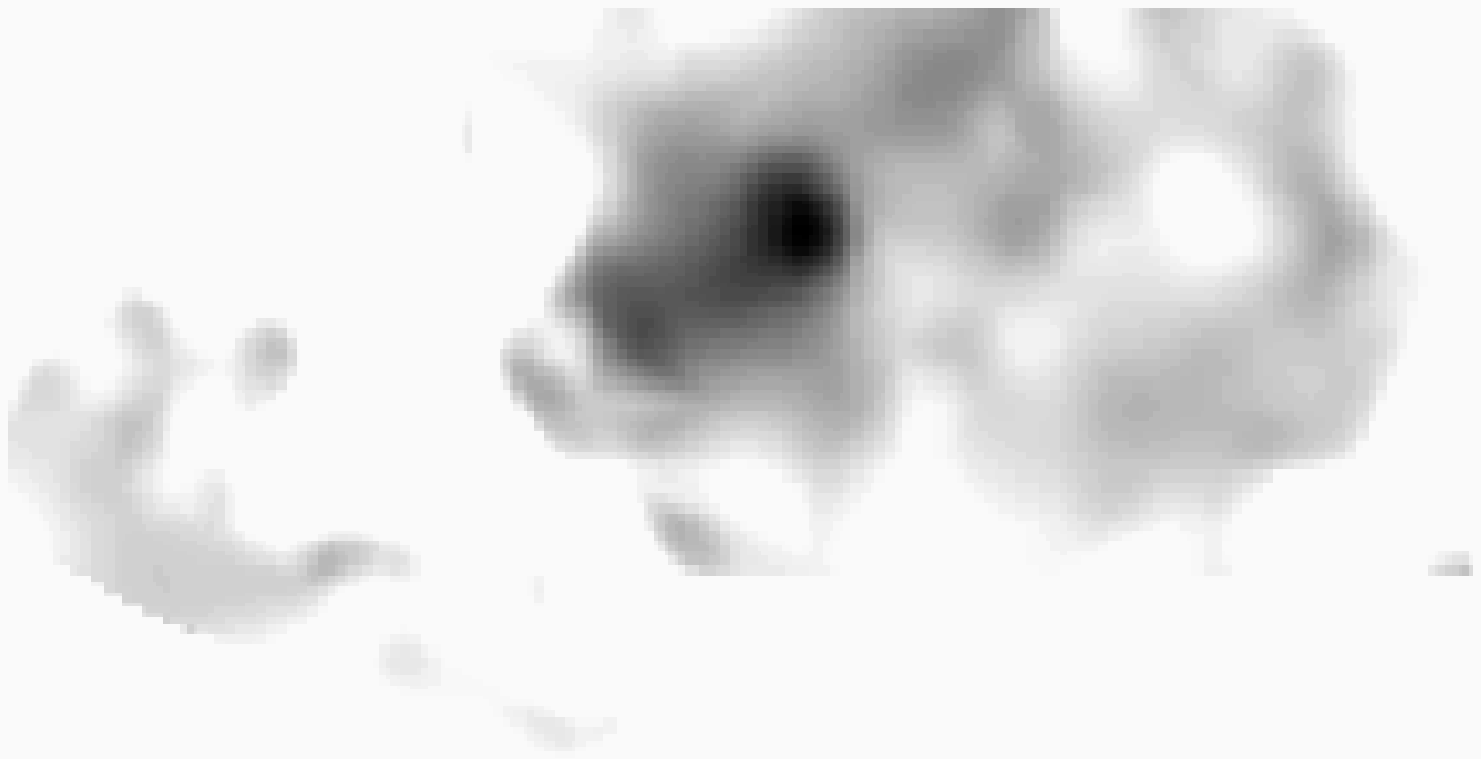}
\caption{\small Top: H-alpha contours overlaid on the soft X-ray greyscale.
Bottom: H-alpha greyscale; field of view is cut off at north boundary.
Data from Colina et al. (2003). }
\end{figure}

\begin{figure}
\epsscale{1.0}
\plotone{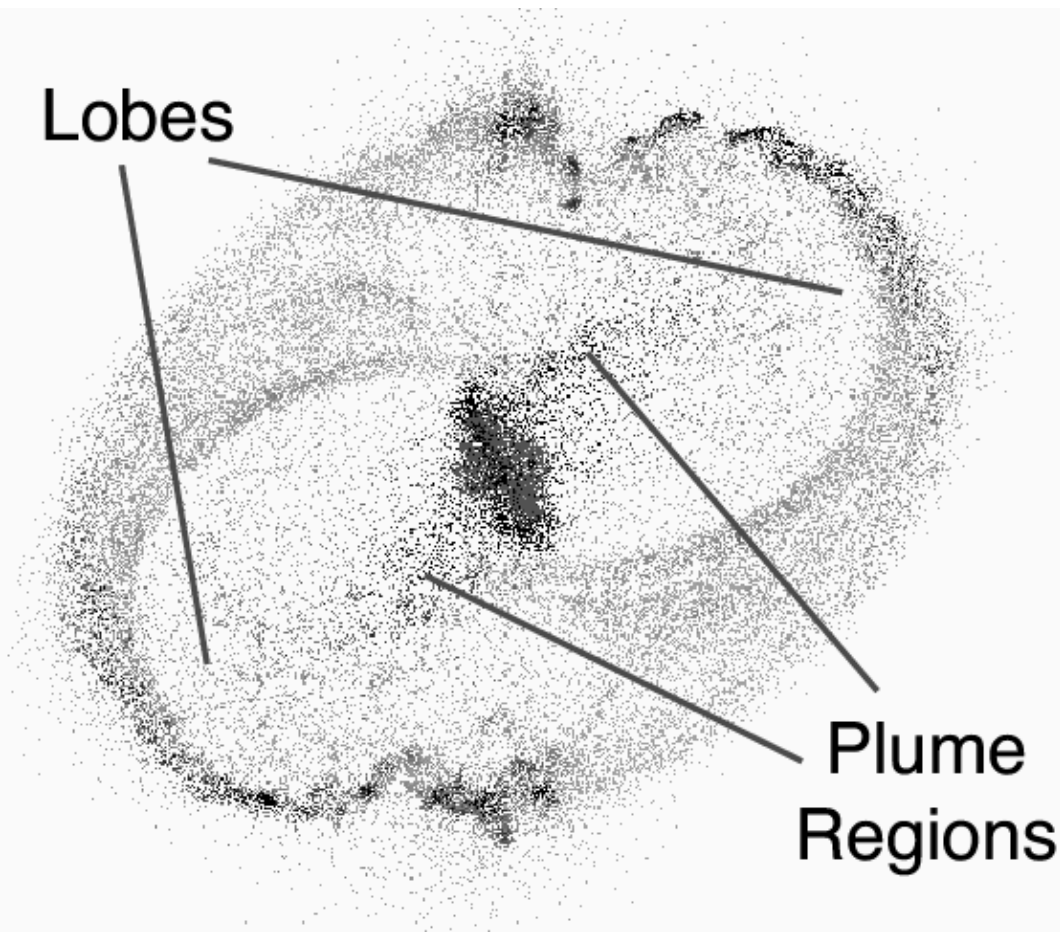}
\caption{\small Dynamical simulation of the Arp 220 merger.  Image shows 
the gas density distribution at a time of $1.5 \times 10^{8}$~yr after 
closest approach and $1.5 \times 10^{7}$~yr after the formation of the 
dense central stellar region. Density is indicated by intensity in the 
paper version, and by color in the electronic version (red represents 
the densest gas and blue the least dense). The system is viewed from 
an angle chosen to match the Arp 220 observations.  The regions 
corresponding to the lobes and the plumes are indicated.  The entire 
gaseous component is shown here, which includes the hot gas seen in 
the X-ray observations, as well as the cool, neutral gas observed 
in H I. The lobe regions are bounded by the arcs of relatively dense 
gas to the far left and far right edges of the distribution.  Much 
of the gas interior to these arcs is falling back toward the nuclear 
region after having been flung to large distances during the 
collision.  The arcs themselves are the remains of large density 
perturbations triggered by the collision dynamics, and are the sites 
of strong shocks.  Material falling into the lobe regions from these 
arcs would have been heated by the large-scale shock waves.  The 
infalling material includes the central plume-like regions along 
the lower-left to upper-right diagonal; this gas could be countered 
by strong outflows from a central starburst in the real system.  
Outside of the lobes, low density gas can be found along the 
upper-left to lower-right diagonal, possibly corresponding to the 
cool, neutral gas seen in H I observations.}

\end{figure}

\begin{figure}
\epsscale{1.0}
\plotone{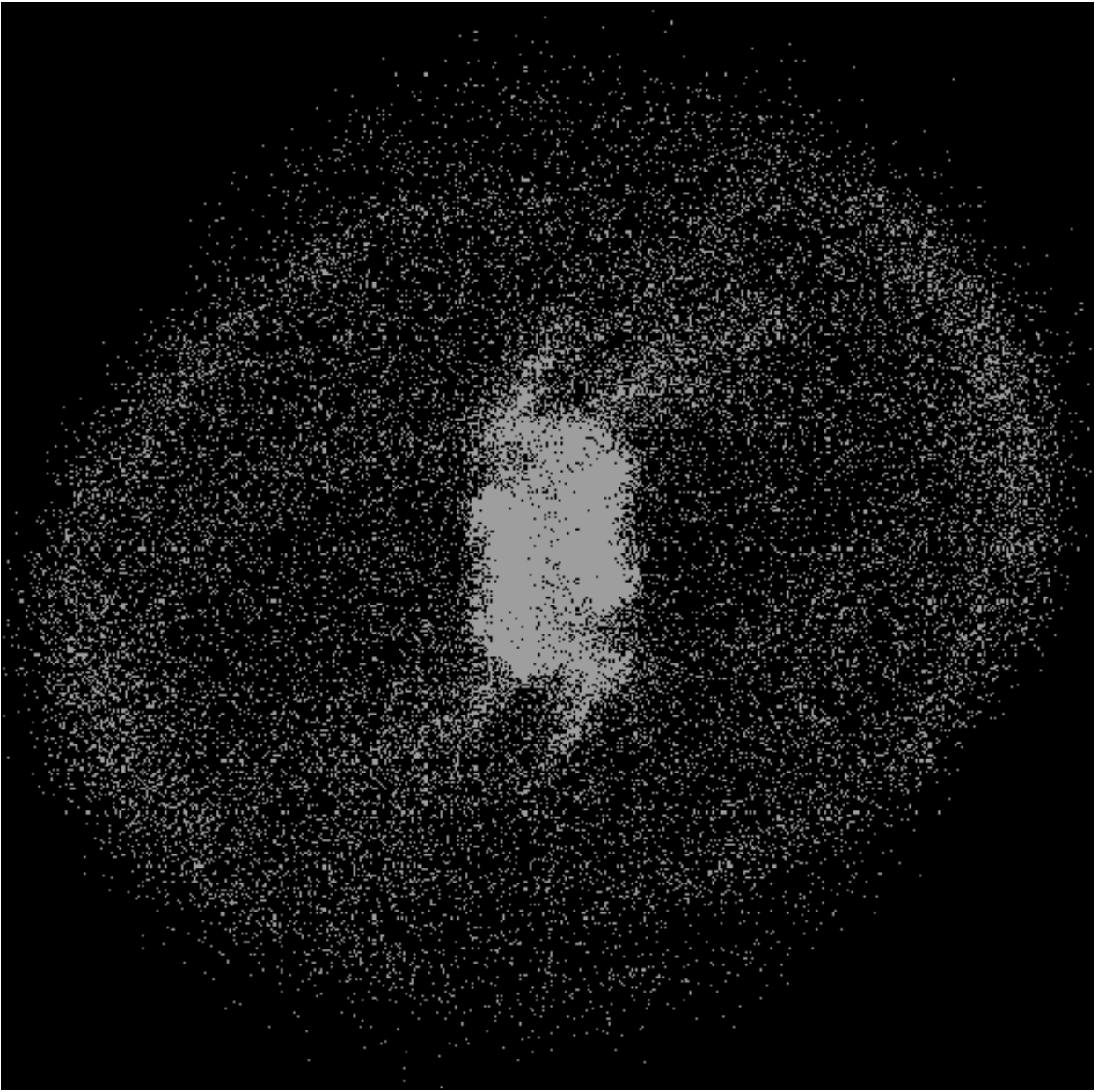}
\caption{\small Dynamical simulation of the Arp 220 merger. Image shows 
the total star density distribution; details as for Fig. 10.}
\end{figure}


\end{document}